\documentclass[12pt]{elsarticle}
\usepackage{amssymb}
\usepackage{lineno}
\usepackage{graphicx}
\usepackage{color}
\usepackage{amssymb,amsfonts,amsmath}
\usepackage{natbib}
\journal{CNSNS}
\begin{document}
\begin{frontmatter}
\title{Ginzburg-Landau approximation for self-sustained oscillators weakly coupled on complex directed graphs}

\author[fisica,infn,CSDC]{Francesca Di Patti}
\author[fisica,infn,CSDC]{Duccio Fanelli}
\author[spagna]{Filippo Miele} 
\author[Namur]{Timoteo Carletti}

\address[fisica]{Dipartimento di Fisica e Astronomia, Universit\`{a} degli Studi di Firenze,  via G. Sansone 1, 50019 Sesto Fiorentino, Italia}

\address[infn]{INFN Sezione di Firenze, via G. Sansone 1, 50019 Sesto Fiorentino, Italia}

\address[CSDC]{CSDC, via G. Sansone 1, 50019 Sesto Fiorentino, Italia}

\address[spagna]{Spanish National Research Council (IDAEA-CSIC), E-08034 Barcelona, Spain}

\address[Namur]{Department of Mathematics and Namur Center for Complex Systems - naXys, University of Namur, rempart de la Vierge 8, B 5000 Namur, Belgium}

\begin{abstract}
A normal form approximation for the evolution of a reaction-diffusion system hosted on a directed graph is derived, in the vicinity of a supercritical Hopf bifurcation. Weak diffusive couplings are assumed to hold between adjacent nodes. Under this working assumption, a Complex Ginzburg-Landau equation (CGLE) is obtained, whose coefficients depend on the parameters of the model and the topological characteristics of the underlying network. The CGLE enables one to probe the stability of the synchronous oscillating solution, as displayed by the reaction-diffusion system above Hopf bifurcation. More specifically, conditions can be worked out for the onset of the symmetry breaking instability that eventually destroys the uniform oscillatory state.  Numerical tests performed for the Brusselator model confirm the validity of the proposed theoretical scheme. Patterns recorded for the CGLE  resemble closely those recovered upon integration of the original 
Brussellator dynamics.
\end{abstract}


\end{frontmatter}
\section{Introduction}

Many real-life phenomena can be ultimately described in terms of mutually interacting entities which can occasionally give rise to collective behaviors \cite{murray2, ball}. The emerging patterns may play important functional roles, as it is for instance the case for biochemical processes \cite{karsentil} and ecological applications \cite{tongway, buhl, bonabeau}. Synchrony is among the most striking example of self-organized dynamics \cite{pikovsky}. It is encountered in a wide gallery of natural systems,  think to the beating of the heart \cite{glass} and the firing of the firefly  \cite{buck}. It also plays a role of paramount importance for the correct functioning of man-designed technology, as e.g. in power grids \cite{rohden, bulloa}. 

Synchronization of self-sustained oscillations is a particularly rich field of investigation.  Mathematically,  self-sustained oscillations correspond to stable limit cycles in the state space of an autonomous continuous-time dynamical system. Oscillators can be embedded in continuum space, being subject to diffusive couplings. The ensuing reaction-diffusion system  displays synchronous oscillations, which, under specific conditions, prove robust to external perturbations. Each oscillator can alternatively occupy a node of a complex network  \cite{arenas2008, osipov, barrat, latora}, a generalization that opens up the perspective to tackle a a large plethora of  problems that deal with a discrete and heterogeneous hosting support  \cite{latora, boccaletti, newmanBook}. 

For continuous reaction-diffusion systems near the supercritical Hopf bifurcation, one can obtain a simplified, normal form description of the dynamics in terms of an associated Ginzburg-Landau equation (CGLE) \cite{kuramotoBook, cross, pikovsky}.  This is a reduced picture which provides an accurate representation of the original model, while allowing for analytical progress to be made. From a  more general perspective, it is also important to classify minimal, though effective descriptive frameworks that could guide in the search for universal traits that happen to be shared by multispecies  models, besides their intrinsic degree of inherent specificity. Simplified schemes that exemplify an exact underlying dynamics can be effectively employed to shed light on the onset of spatio-temporal chaos  and propagation of nonlinear waves.  Externally imposed non homogeneous perturbation can, for instance, break the synchrony of the oscillations as displayed by the original reaction-diffusion system, or equivalently its CGLE analogue, so materializing in a colorful density patterns that sustain spatio-temporal propagation. The formal link between continuous reaction diffusion-systems and the CGLE was established in the pioneering work by Kuramoto \cite{kuramotoBook}, exploiting a multiple-scale perturbative analysis. More recently the analysis has been extended by Nakao \cite{nakao2014} to the relevant setting where the reaction-diffusion system is made to evolve on a symmetric, hence undirected, network. As remarked in \cite{asllani14}, directionality matters and can seed the emergence of non trivial collective dynamics which cannot manifest when the scrutinized system is  made to evolve on a symmetric discrete support. Motivated by this finding, we considered in \cite{contemori} the dynamics of a reaction-diffusion system defined on a directed graph which displays a stable fixed point and obtained an effective description for the evolution mode triggered unstable, just above the threshold of criticality. The analysis exploits a multiple time-scale analysis and eventually yields a Stuart-Landau for the amplitude of the unstable mode, whose complex coefficients reflect the topology of the network, the factual drive to the instability.

In this paper we aim at following the similar strategy of \cite{contemori} to derive an approximate equation for the evolution of a reaction diffusion system on a directed (and balanced) graph, in the vicinity of a supercritical Hopf bifurcation. As a matter of fact we will assume weak the strength of the coupling that links adjacent nodes. In doing so, we will generalize the work of Nakao \cite{nakao2014} to the interesting setting where asymmetry in the couplings needs to be accommodated for and, at the same time, reformulate the classical work of Kuramoto \cite{kuramotoBook}, on a discrete spatial backing.  To anticipate our findings, and at variance with the analysis reported in \cite{contemori}, we will finally obtain a CGLE as a minimal description for the dynamics of the self-sustained oscillators coupled on a complex and asymmetric graph. The obtained CGLE enables one to analytically probe the stability of the synchronous uniform state, as displayed by the reaction-diffusion system. Specifically, it allows to determine the parameters setting that instigates a symmetry breaking transition to non-uniform patterns. Numerical tests made for the Brusselator model, here assumed as a reference model for its pedagogical interests, will confirm the adequacy of the proposed approximate scheme.
\section{Diffusive  oscillators on networks}
We will here consider a generic two dimensional reaction-diffusion system and label with ${\bf x}_j(t)=(\phi_j, \psi_j)^T$ for $j=1, \ldots, N$ the two-dimensional real vector of the concentrations. The index $j$ refers to the node of the network to which the selected components refer to. $N$ stands for the size of the network, i.e. the total number of nodes. The only further assumption that we shall make is the existence of self-sustained oscillations for the system under scrutiny and for this reason we will point to ${\bf x}_j(t)$ as to the oscillators' variables. The dynamics of the system is hence described by the following differential equation
\begin{equation}\label{eq:systemRD} 
\dot{\bf x}_j = {\bf F} ({\bf x}_j , \boldsymbol{\mu}) +{\bf  D} \sum_{k=1}^N \Delta_{jk}{\bf x}_k
\end{equation}
where the two-dimensional nonlinear function $ {\bf F} $ specifies the reaction terms: it depends on the local  concentrations of the species ${\bf x}_j$ and on $\boldsymbol{\mu}$ which is a vector of arbitrary dimension that  gather together the parameters of the model. The second term represents the diffusive coupling: ${\bf D}= K \text{diag}(D_{\phi},D_{\psi})$ denotes the diagonal matrix of the diffusion coefficients. $K$ is a constant parameter that set the strength of the coupling. As we will make clear in the following the perturbative analysis that we shall develop, hold for $K<<1$. In equation \eqref{eq:systemRD}, 
${ \bf \Delta}$ is the Laplacian matrix whose elements read $\Delta_{ij}=A_{ij}-\delta_{ij}k_i$. Here we focus on directed networks, thus the adjacency matrix ${\bf A}$ is not symmetric.  Using standard notations, $A_{ij}=1$ if a link exists that goes from  node $i$ to node  $j$. Otherwise, $A_{ij}=0$. $k_i$ is the number of outgoing edges from node $i$, $\delta_{ij}$ is the Kronecker's delta. We assume that system \eqref{eq:systemRD} admits a homogeneous equilibrium point that we here label ${\bf x}^*=(\phi^*, \psi^*)^T$. This request implies dealing with a balanced network, namely a network where the outgoing and incoming connectivities are equal. 

We additionally require that ${\bf x}^*$ undergoes a Hopf bifurcation for $\boldsymbol{\mu} =\boldsymbol{\mu}_0$. Accordingly, the Jacobian matrix associated to system  \eqref{eq:systemRD}  has a pair of imaginary eigenvalues $\pm i \omega_0$. Slightly above the supercritical Hopf bifurcation, ${\bf x}^*$ becomes unstable, the reaction-diffusion system admits an time dependent homogeneous solution. This is the uniform state obtained by replicating on each node of the network and in complete synchrony, the  limit cycle displayed by the system in its a-spatial limit ($K=0$). The spatial coupling, sensitive to tiny non homogeneities, which configure as injected perturbation, can eventually destabilized the uniform synchronous equilibrium.  When diffusion is small ($K=\epsilon^2 <<1$), the method of multiple timescales \cite{kuramotoBook} constitutes a viable strategy to characterize the nonlinear evolution of the perturbation and hence elaborate on the stability of the time-dependent uniform periodic solution. 

To this end, inspired by  the analysis carried out in \cite{nakao2014}, we introduce small inhomogeneous perturbations, $\delta \phi_j$ and $\delta \psi_j$, to the uniform equilibrium point, namely $(\phi_j, \psi_j)=(\phi^*, \psi^*)+(\delta \phi_j, \delta \psi_j)$ for $j=1, \ldots, N$. We then substitute this ansatz into equations \eqref{eq:systemRD}. Performing a Taylor expansion of the resulting system and packing  $\delta \phi_j$ and $\delta \psi_j$ into the column vector ${\bf u}_j=\left ( \delta \phi_j, \delta \psi_j \right )^T$,  one ends up with  the following equation for the time evolution of  ${\bf u}_j$: 
\begin{equation}\label{eq:equation_u}
\frac{\partial}{dt} { \bf u}_j= {\bf  L  u}_j  +  {\bf D} \sum_{k=1}^N \Delta_{jk} {\bf u}_k+  \mathcal{M}  { \bf u}_j {\bf u}_j +  \mathcal{N}   {\bf u}_j  {\bf u}_j  {\bf u}_j + \ldots
\end{equation}
where ${\bf L}$ is the Jacobian matrix evaluated at the steady state ${\bf x}^*$. Adopting the same symbolic notations of \cite{kuramotoBook}, $\mathcal{M}  {\bf u}_j  {\bf u}_j  $ and $ \mathcal{N}  {\bf u}_j  {\bf u}_j  {\bf u}_j $ denote two-dimensional vectors whose components respectively read
\begin{equation*}
\begin{aligned}
\left ( \mathcal{M} {\bf u}_j  {\bf u}_j  \right  )_l &=\frac{1}{2 !}  \sum_{m,n=1}^2 \frac{\partial^2 F_l ( {\bf x}^*, \boldsymbol{\mu})}{\partial x_n \partial x_m}  ( {\bf u}_j )_m    ( {\bf u}_j )_n \\
\left ( \mathcal{N}  {\bf u}_j  {\bf u}_j  {\bf u}_l   \right  )_l & =\frac{1}{3 !}  \sum_{m,n,p=1}^2 \frac{\partial^3 F_l ( {\bf x}^*, \boldsymbol{\mu})}{\partial x_n \partial x_m \partial x_p}  ( {\bf u}_j )_m   ( {\bf u}_j )_n ( {\bf u}_j )_p 
\end{aligned}
\end{equation*}
for $l=1,2$. 

Setting the model above the supercritical Hopf bifurcation, translates into redefining  $\boldsymbol{ \mu}$ by means of a small parameter $\epsilon$ as $\boldsymbol{ \mu}=\boldsymbol{ \mu}_0+ \epsilon^2 \boldsymbol{ \mu}_1$, where  $\boldsymbol{ \mu}_1$ is order one. To follow the nonlinear evolution of the amplitude of the most unstable mode, it is appropriate to introduce a slow time variable $\tau= \epsilon^2 t$ and, accordingly, to  rewrite the total derivative with respect to the original time $t$: 
\begin{equation}\label{eq:expansionTime}
\frac{d}{d t} \longrightarrow \frac{\partial}{\partial t} + \epsilon^2 \frac{\partial}{\partial \tau} \quad .
\end{equation}
As a key assumption,  already alluded to in the preceding discussion, we set $K=\epsilon^2$: in other words we suppose that the impact of diffusion is of the same order as the deviation from the bifurcation point.  
Moreover, near criticality, the matrix ${\bf L}$ and the operators $\mathcal{M}$ and $\mathcal{N}$ may be expanded in powers of $\epsilon^2$ 
\begin{equation}\label{eq:operatorExpansions}
\begin{aligned}
\bf
L&=& {\bf L}_0 + \epsilon^2 {\bf L}_1+\ldots \\
\mathcal{M}&=& \mathcal{M}_0 + \epsilon^2 \mathcal{M}_1+\ldots \\
\mathcal{N}&=& N_0 + \epsilon^2 \mathcal{N}_1+\ldots
\end{aligned}
\end{equation}
We further assume that ${\bf u}$ can be expressed as a series, function of both $t$ and $\tau$:
\begin{equation}\label{eq:expansionU}
{\bf u}_j(t) = \epsilon {\bf u}_j^{(1)}(t,\tau)+\epsilon^2 {\bf u}_j^{(2)}(t,\tau)+ \ldots
\end{equation}
Inserting Eqs. \eqref{eq:expansionTime},  \eqref{eq:operatorExpansions} and \eqref{eq:expansionU} into Eq. \eqref{eq:equation_u} we get:  
\begin{multline*}
\hspace{-0.25cm}\left(\frac{\partial}{\partial t  } \mathbb{I}_{2} +\epsilon^2 \frac{\partial }{\partial \tau}\mathbb{I}_{2}  - {\bf L}_0  -\epsilon^2 {\bf L} _1 - \ldots  \right ) (\epsilon {\bf u}_j^{(1)} +  \epsilon ^2  {\bf u}_j^{(2)} +\ldots) \\
- \epsilon^2 { \bf D} \sum_{k=1}^N \Delta_{jk} (\epsilon {\bf u}_k^{(1)} + \epsilon ^2  {\bf u}_k^{(2)} +\ldots)
= \epsilon^2 \mathcal{M}_0 {\bf u}_j^{(1)} {\bf u}_j^{(1)} + 
\epsilon^3 (2 \mathcal{M}_0 {\bf u}_j^{(1)} {\bf u}_j^{(2)}  \\
+\mathcal{N}_0 {\bf u}_j^{(1)} {\bf u}_j^{(1)} {\bf u}_j^{(1)}) + \mathcal{O} (\epsilon^4) 
\end{multline*}
Collecting together terms of the same order in $\epsilon$ returns the following set of equations 
\begin{equation}\label{eq:terms}
\left ( \frac{\partial}{\partial t }  \mathbb{I}_{2} - {\bf L}_0   \right ) {\bf u}_j^{(\nu)}={\bf B}_j^{(\nu)}  
\end{equation}
for $\nu=1,2,3...$. The existence of a nontrivial solution for the aforementioned linear systems is guaranteed by the Fredholm theorem \cite{fredholm}. As we shall discuss in a moment, the solvability condition (see \ref{sec:solvability}) is directly satisfied for $\nu=1$ and $\nu=2$, while it must be explicitly imposed for $\nu=3$. 
 
Let us start by focusing  on $\nu=1$.  The corresponding right hand side of system (\ref{eq:terms}) is ${\bf B}_j^{(1)}   =  0$. It is easy to see that the analytical solution is
\begin{equation}
\label{eq:u1_W} 
{\bf u}_j^{(1)}(t, \tau) =   W_j(\tau) {\bf U}_0 e^{ i \omega_0 t} + c. c. 
\end{equation}
where ${\bf U}_0$ the right eigenvector of $ {\bf L}$ corresponding to the eigenvalue $i \omega_0$. The complex variable $W_j(\tau)$ is the amplitude of the perturbation. Its dynamics is supposed to be slow, i.e. ruled by the rescaled time $\tau$. As we shall see, by imposing the solvability condition for $\nu=3$ will return an equation for the time evolution of $W_j(\tau)$. 

Moving to $\nu=2$, we find ${\bf B}_j^{(2)}  =  \mathcal{M}_0 {\bf u}_j^{(1)} {\bf u}_j^{(1)}$ and we try to cast the solution of the linear system in the form  
\begin{equation*}
{\bf u}_j^{(2)}= W_j^2 {\bf V}_2  e^{2 i \omega_0 t } + \gamma {\bf u}_j^{(1)} + \vert W_j \vert ^2 {\bf V}_0 \qquad .
\end{equation*}
Inserting this ansatz into \eqref{eq:terms} and grouping together  terms that do not depend on $t$, one finds $ {\bf V}_0 = -2  {\bf L}_0^{-1} \mathcal{M}_0 {\bf U}_0 \bar{{\bf U}} _0$ where the bar stands for the complex conjugate.  In the same way, collecting  terms proportional to $e^{2 i \omega_0 t } $  yields  $ {\bf V}_{2} = \left ( 2 i \omega_0 \mathbb{I}_{2} -L_0 \right ) ^{-1} \mathcal{M}_0 {\bf U}_0 {\bf U}_0$. 

The third constant term appearing in  Eq. (\ref{eq:terms}) is explicitly given by
${\bf B}_j^{(3)}   =   \left ( {\bf L} _1 -\frac{d}{d \tau } \mathbb{I}_{2}  \right ) {\bf u}_j^{(1)}  $ $ +  { \bf D} \sum_{k=1}^N \Delta_{jk} {\bf u}_k^{(1)} + 2 \mathcal{M}_0 {\bf u}_j^{(1)} {\bf u}_j^{(2)} + \mathcal{N}_0 {\bf u}_j^{(1)} {\bf u}_j^{(1)} {\bf u}_j^{(1)}$. By imposing the solvability condition, one eventually obtains the following CGL equation for $W_j(\tau)$ 
\begin{equation}\label{eq:GLE}
\frac{d}{d \tau} W_j(\tau) = \sigma W_j -  g \vert W_j \vert ^2 W_j + d \sum_{k=1}^N \Delta_{jk} W_k
\end{equation}
where $\sigma= \sigma_{Re}+ i \sigma_{Im} =({\bf U}_0^*)^\dagger  {\bf L}_1 {\bf U}_0 $,  $ g= g_{Re} + i g_{Im} = $ $ -({\bf U}_0^*)^\dagger  \left [ 2 \mathcal{M}_0 {\bf V}_2 \bar{\bf U}_0  \right . $ $ \left . +2 \mathcal{M}_0   {\bf V}_0 {\bf U}_0 +  3 \mathcal{N}_0  {\bf U}_0   {\bf U}_0    \bar{\bf U}_0   \right ]$ and $d=d_{Re}+ i d_{Im}= ({\bf U}_0^*)^\dagger {\bf D} {\bf U}_0 $ are complex numbers. 

Using the following scale transformation
\begin{equation*}
\frac{\tau}{\sigma_{Re}}  \rightarrow \tau  \qquad \frac{d_{Re}}{\sigma_{Re}} \Delta_{ij}  \rightarrow  \Delta_{ij} \qquad \sqrt{\frac{g_{Re}}{\sigma_{Re}}} W_j   \rightarrow W_j  \quad, 
\end{equation*}
introducing three new coefficients
\begin{equation*}
c_0 = \frac{\sigma_{Im}}{\sigma_{Re}} \qquad c_1=\frac{d_{Im}}{d_{Re}} \qquad c_2 =  \frac{g_{Im}}{g_{Re}}
\end{equation*}
and making the further transformation  $W_j  \rightarrow W_j \exp(i c_o \tau)$, we find the following reduced CGLE 
\begin{equation}\label{eq:GLsimple}
\frac{d}{d \tau} W_j =  W_j - (1+i c_2) \vert W_j \vert ^2 W_j + (1+i c_1) \sum_k \Delta_{jk} W_k
\end{equation}
for the complex amplitude $W_j$ of the $j$-th oscillator.  

Equation \eqref{eq:GLsimple} displays a uniformly synchronized solution given by 
\begin{equation}\label{eq:limitCycle}
W_j(\tau)\equiv W^{LC}=  e^{- i c_ 2 \tau} 
\end{equation}
for $j=1,\ldots, N$. In this respect it provides a local approximation for the dynamics of the original reaction-diffusion system near the Hopf bifurcation, namely when self-organized oscillations rise. The dynamics of the ensemble made of $N$ oscillators is sensitive to the coupling imposed, this latter being encoded in the Laplacian operator. Mutual interferences among oscillators, as mediated by the weak diffusive coupling here at play, can disrupt the coherency of the synchronous state, making the system to evolve towards a different attractor. The stability of the synchronized dynamics can be assessed via a linear stability analysis which is carried out for CGLE, and whose conclusion can be readily translated to the original reaction-diffusion framework. Having obtained a universal description of the dynamics of the system in terms of a normal mode representation allows one to obtain general conditions of the onset of the instability, which prescind the specific context of analysis. 

To investigate the linear stability of solution \eqref{eq:limitCycle}, we briefly review the method described in \cite{nakaoCL, nakao2014} and point to the specific aspects that relate to the directed nature of the couplings, as discussed in \cite{dipattiBFreti}. Notice that in this latter paper, the CGLE is assumed as the reference model, while here it is obtained as a local approximation of a generic reaction-diffusion system defined on a directed network, under weak diffusive couplings. 

To carry out the linear stability calculation, we substitute the following  amplitude and phase perturbation \cite{nakaoCL, nakao2014}
\begin{equation*}
W_j (\tau) =W^{LC}(\tau) (1 + \rho_j(\tau))e^{i \theta_j( \tau)}
\end{equation*}
into Eq. \eqref{eq:GLsimple} and we linearize around $\rho_j=0$ and $\theta_j=0$ obtaining 
\begin{equation}\label{eq:linearization}
\left (
\begin{matrix}
\dot{\rho}_j\\
\dot{\theta}_j
\end{matrix}
\right)
= 
\left (
\begin{matrix}
-2 & 0 \\
-2 c_2 & 0
\end{matrix}
\right)
\left (
 \begin{matrix}
\rho_j\\
\theta_j
\end{matrix}
\right)
+ \sum_{k=1}^N \Delta_{jk}
 \left (
\begin{matrix}
1 & -c_1 \\
c_1 & 1
\end{matrix}
\right)
\left (
 \begin{matrix}
\rho_k\\
\theta_k
\end{matrix}
\right) \quad .
\end{equation}
Using the Laplacian eigenvalues and eigenvectors that, by definition, satisfy  ${\bf \Delta } {\bf \Phi}^{(\alpha)} $ $=  \Lambda^{(\alpha)} {\bf \Phi}^{(\alpha)}$ for $\alpha=1, \dots, N$, the inhomogeneous perturbations $\rho_j$ and $\theta_j$ can be expanded  as 
\begin{equation}\label{eq:expansionEigenvectorsDelta}
\begin{pmatrix}
\rho_j \\ \theta_j
\end{pmatrix}= \sum_{\alpha=1}^N
\begin{pmatrix}
\rho^{(\alpha)} \\ \theta^{(\alpha)}
\end{pmatrix}
e^{\lambda^{(\alpha)} \tau} \Phi_j^{(\alpha)}
\end{equation}
where $\rho^{(\alpha)} $  and $\theta^{(\alpha)}$ denote the expansion coefficients, while $\lambda^{(\alpha)}$ controls the exponential growth of the $\alpha$-th mode. If the real part of $\lambda^{(\alpha)}$ ($\lambda^{(\alpha)}_{Re}$) is negative, perturbations shrink to zero, otherwise they grow exponentially. Plugging expansion \eqref{eq:expansionEigenvectorsDelta} into equation \eqref{eq:linearization} yields  the following condition 
\begin{equation}\label{eq:relDisp}
\text{det}  \left (  {\bf J}_{\alpha}  - \lambda^{(\alpha)} \mathbb{I}_{2}  \right  ) =0
\end{equation}
with 
\begin{equation*}
{\bf J}_{\alpha} = \left (
\begin{matrix}
-2 +\Lambda^{(\alpha)} & -c_1 \Lambda^{(\alpha)}\\
-2 c_2 + c_1 \Lambda^{(\alpha)} & \Lambda^{(\alpha)}
\end{matrix}
\right ) \qquad .
\end{equation*}
Constrain (\ref{eq:relDisp}) returns an equation for  $\lambda^{(\alpha)}$ as a function of $\Lambda^{(\alpha)}$ known as the dispersion relation. For a symmetric undirected network, as the one explored by Nakao \cite{nakaoCL, nakao2014}, the eigenvalues $\Lambda^{(\alpha)}$ of the Laplacian operators are real and this observation translates in simple condition for the instability to develop. A straightforward calculation, allows one to conclude that the perturbation can grow, and so destroy the synchrony of the system, provided $1+c_1 c_2 <0$. When  the underlying graph of coupling is instead directed, the spectrum of the Laplacian operator is complex and the instability can set in also when $1+c_1 c_2 >0$. For this reason, the generalized class of instability that is encountered when the system is confined on a asymmetric support is referred to as to topology driven. The conditions which result in $\lambda^{(\alpha)}_{Re}>0$ when the network of couplings is directed, or, equivalently, when $\Lambda^{(\alpha)}$ is complex ($\Lambda^{(\alpha)}=\Lambda_{Re}^{(\alpha)} + i \Lambda_{Im}^{(\alpha)}$), have been worked out in  \cite{dipattiBFreti}, building on the recipe outlined in \cite{asllani14}. The conclusion of \cite{dipattiBFreti} are here briefly reviewed for for the sake of completeness. Indeed, $\lambda_{Re}^{(\alpha)}>0$ if 
\begin{equation}\label{eq:inequality}
S_2(\Lambda_{Re}^{(\alpha)}) \leqslant S_1(\Lambda_{Re}^{(\alpha)}) \left [ \Lambda_{Im}^{(\alpha)} \right ] ^2
\end{equation}
where 
\begin{equation*}
\begin{aligned}
S_2(\Lambda_{Re}^{\alpha}) &=C_{2,4}(\Lambda_{Re}^{(\alpha)})^4-C_{2,3}(\Lambda_{Re}^{(\alpha)})^3+C_{2,2}(\Lambda_{Re}^{(\alpha)})^2\\
& \phantom{= }-C_{2,1}\Lambda_{Re}^{(\alpha)} \\
S_1(\Lambda_{Re}^{\alpha}) &=C_{1,2}(\Lambda_{Re}^{\alpha})^2 -C_{1,1}\Lambda_{Re}^{(\alpha)} +C_{1,0}
\end{aligned}
\end{equation*}
with
\begin{equation*}
\begin{aligned}
C_{2,4}&=1+c_{1}^2  \\
C_{2,3}&=4+2c_{1}c_{2}+2c_{1}^2 \\
C_{2,2}&=5+4c_{1}c_2+c_{1}^2 \\
C_{2,1}&=2+2c_{1}c_2 \\
C_{1,2}&=c_1^4+c_1^2 \\
C_{1,1}&=2c_1^{3}c_2+2c_1^2  \\
C_{1,0}&=c_{1}^{2}(1+c_2^2) \qquad .
\end{aligned}
\end{equation*}
When condition (\ref{eq:inequality}) is met for some eigenvalue $\Lambda^{(\alpha)}$, the limit cycle looses its stability and the system evolves towards a different  attractors, which is non homogeneous in space. 

The next section is aimed at challenging the validity of Eq. (\ref{eq:GLsimple}), as an approximate equation for  the non linear evolution of an injected non homogeneous  perturbation. To anticipate our findings, we will show that the CGLE allows to correctly delineate the region of parameters that yields stable and synchronized self-sustained oscillators, and so accurately mark the transition to the pattern forming domain.  In both cases, i.e. when uniform oscillations or complex patterns in density are displayed, we obtain a satisfying degree of correspondence between the original reaction-diffusion model and its corresponding CGLE. To perform the numerical tests we will make use of the Brusselator model, as a reference case study. 
\section{Numerical validation of the theory: the Brusselator model}\label{sec:brusselator}
To challenge the analysis performed in the previous section, we here consider a specific reaction-diffusion system, the Brusselator model, which allows for self-sustained oscillations. The governing equations read:
\begin{equation}\label{eq:RDbrusselator}
\left \{
\begin{aligned}
\dot{\phi_j} &=  A-(B+1)  \phi _j + \phi_j^2  \psi_j + D_{\phi} \sum_{k=1}^N \Delta_{jk} \phi_k \\
\dot{\psi_j} &=  B \phi_j - \phi_j^2  \psi_j + D_{\psi} \sum_{k=1}^N \Delta_{jk} \psi_k 
\end{aligned}\right. 
\end{equation}
where $A$, $B$ stand for non negative constants parameters. This system admits a homogeneous steady state given by $(\phi^*,\psi^*)=(A,B/A)$. By acting on the parameter $B$, while keeping $A$ fixed, one can induce a Hopf bifurcation, which opens up the route to self-sustained oscillations. The bifurcation occurs for $B_c=1+A^2$, as explained in the annexed  \ref{sec:detailsBrusselator}. Above the bifurcation point, the Brusselator dynamics can be approximated by a CGLE. Following the derivation reported in the   \ref{sec:detailsBrusselator}, which builds on the seminal paper by Kuramoto \cite{kuramotoBook}, one ends up with the compact expression for  $c_1$ and $c_2$: 
\begin{equation}\label{eq:c1_c2}
\begin{aligned}
c_1 & =-A\dfrac{D_{\phi}-D_{\psi}}{D_{\phi}+D_{\psi}} \\
c_2 & =\dfrac{4-7A^2+4A^4}{3A(2+A^2)} \qquad .
\end{aligned}
\end{equation}
 
The CGLE is therefore completely characterized and it can be readily employed to asses the stability of the uniform oscillatory state, namely the spatially extended configuration that is obtained as the synchronous replica of $N$ self-sustained oscillators. Fix the values of  $D_{\phi}$ and $D_{\psi}$. Then, the stability of the synchronous state is ultimately controlled by  $A$, the sole parameter that one can freely tune. We will denote with $A_c$ the value of $A$ (if it exists), above which stability is eventually lost. Moreover, we will set the parameters $c_1$ and $c_2$ so as to enforce stability on a symmetric support, namely $1+c_1c_2<0$. 
The instability that materializes for $A>A_c$ is hence indirectly reflecting the asymmetry of the imposed couplings. 

\begin{figure*}[tb]
\begin{tabular}{ccc}
\includegraphics[scale=0.22]{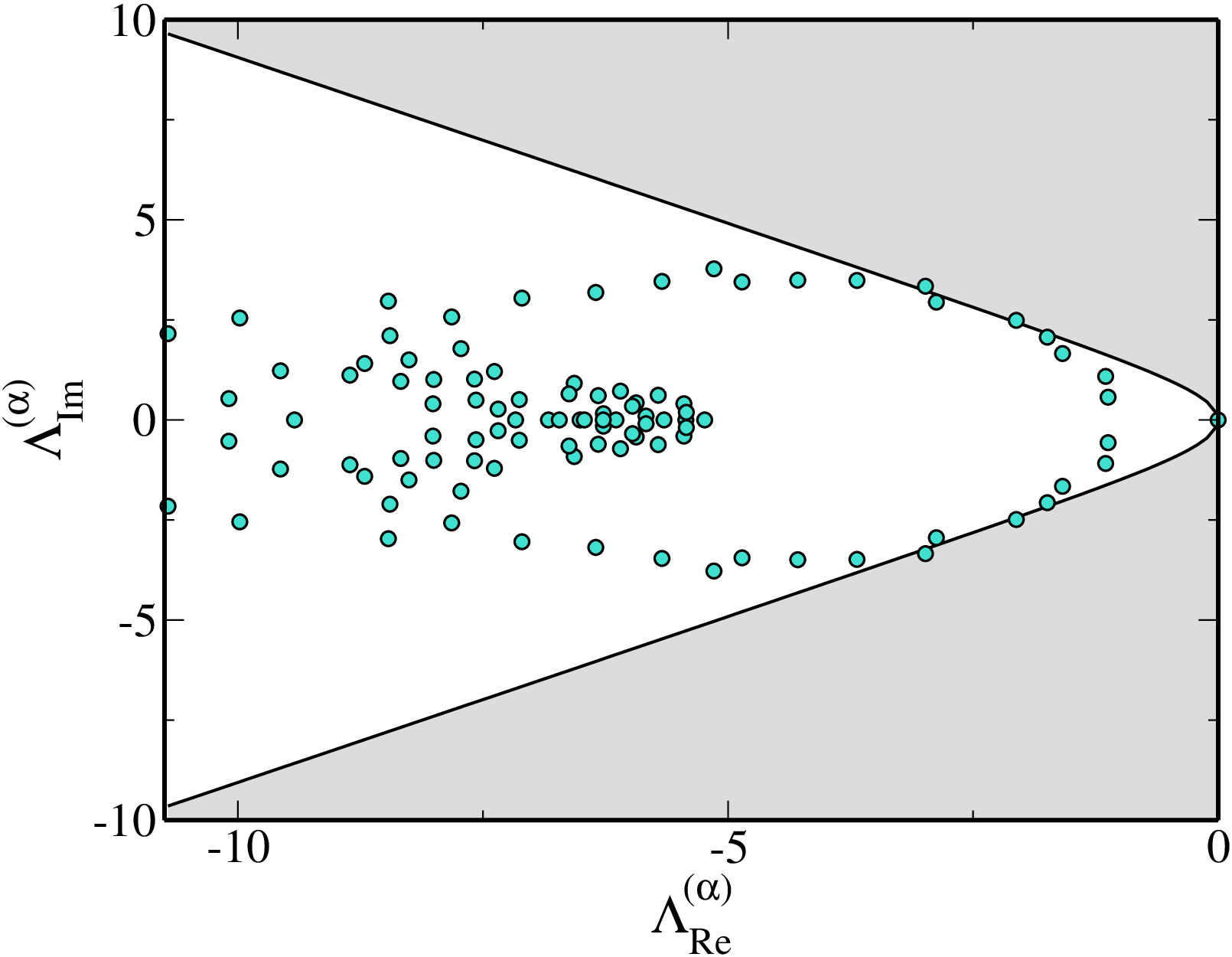} &
 \phantom{c} &
\includegraphics[scale=0.22]{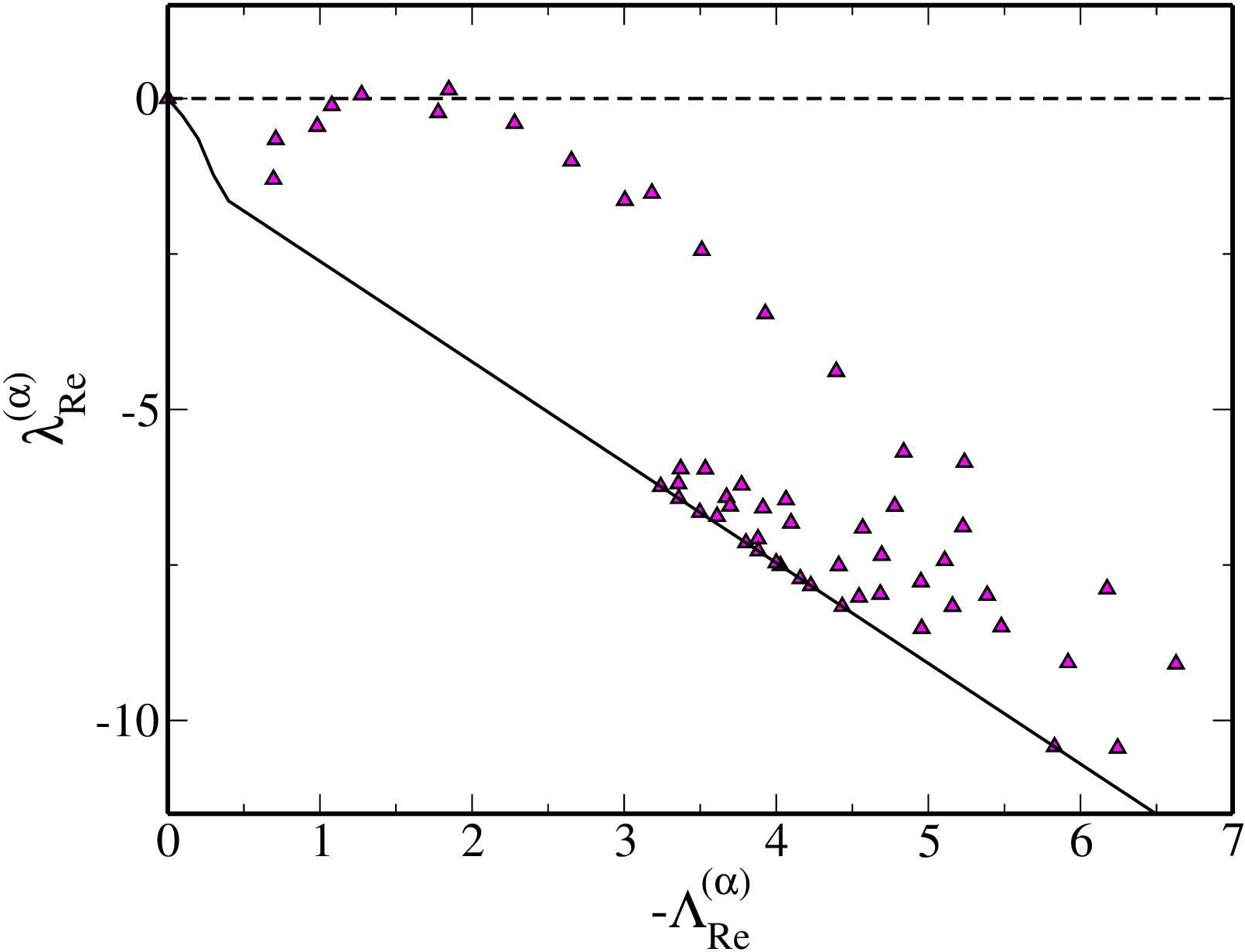}\\
(a) &  \phantom{c} & (b) \\
 \phantom{c} &  \phantom{c}\\
\includegraphics[scale=0.31]{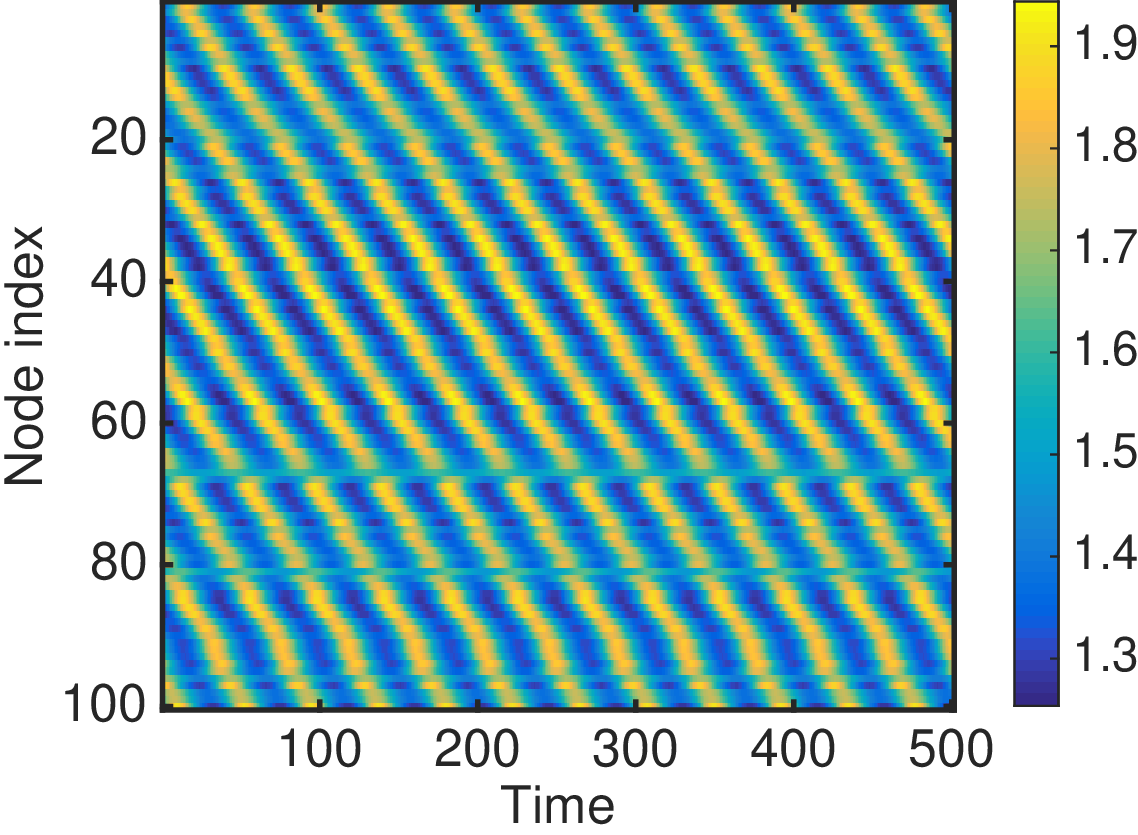} &
 \phantom{c} &
\includegraphics[scale=0.31]{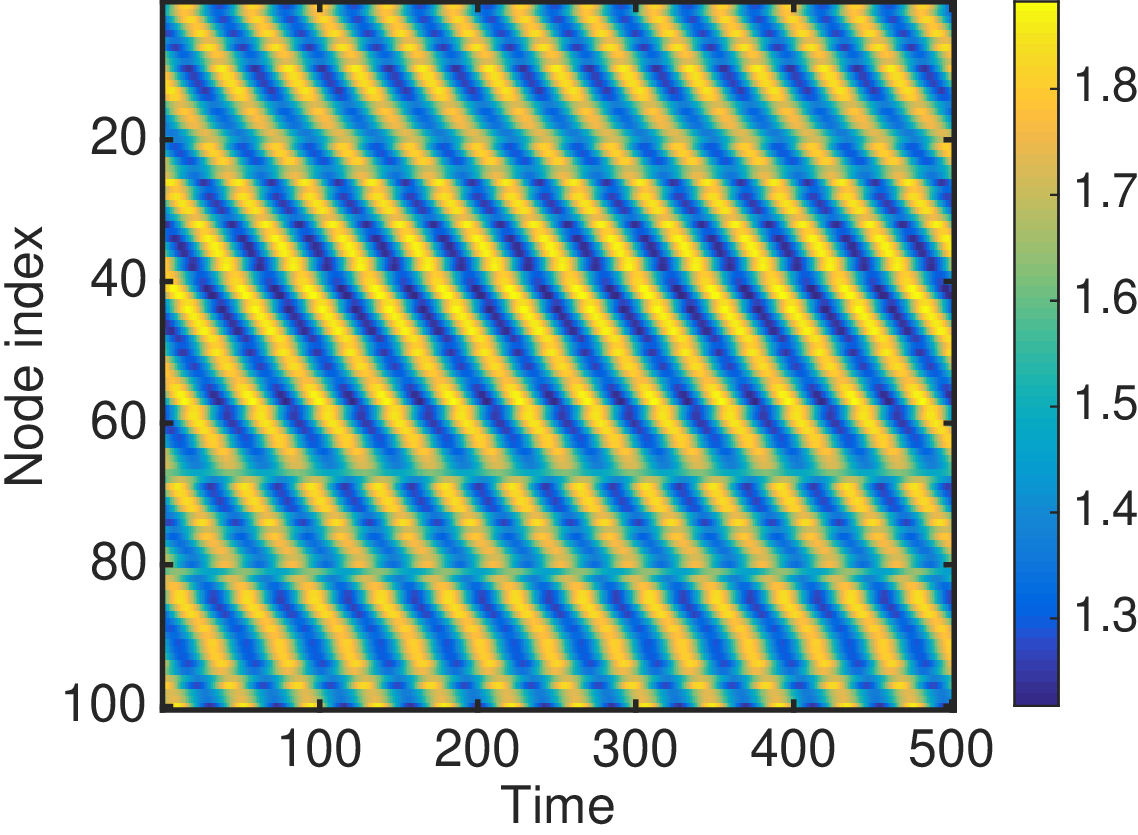}\\
(c) &  \phantom{c} & (d)
\end{tabular}
\caption{Panel (a): eigenvalues of the Laplacian for a balanced Newman-Watts \cite{newman99} network generated with $p=0.27$ and $N=100$ nodes. The algorithm to built the network is detailed in \cite{contemori}. The shaded area marks the instability region for the Brusselator model, as obtained under the CGL normal form representation. Here,
$A=1.55$, $B_c=1+A^2$, $D_{\phi}=0.6$ and $D_{\psi}=4.9$. The perturbative parameter is  $\epsilon=0.07$. For the case at hand, $A_c=1.48$. Panel (b): the real part of the dispersion relation (magenta triangles) for the same choice of network and parameters as in panel (a). The black line originates from the continuous theory. Panel (c): pattern relative to species $\phi$ obtained by numerical integration of system \eqref{eq:RDbrusselator}. Panel (d): pattern relative to species $\phi$  calculated using the CGL approximation.  \label{fig:a155}}
\end{figure*}

The results of the analysis is depicted in Fig. \ref{fig:a155}(a) and (b). The shaded grey area in the plane ($\Lambda_{Re}^{(\alpha)}, \Lambda_{Im}^{(\alpha)}$) delimits the region of instability, as it follows inequality \eqref{eq:inequality}. Symbols represent the eigenvalues of the Laplacian matrix in the complex plane and follow the specific choice of the network operated. As explained in the caption of Fig. \ref{fig:a155}, we here work with a balanced Newman-Watts graph, generated by following the recipe discussed in \cite{contemori}. For the selected value of $A$ (larger than the critical value $A_c=1.48$), two eigenvalues of the Laplacian matrix protrude inside the shaded region, thus triggering the instability. This is indeed a topology driven instability, as it cannot realize if the same choice of parameters is operated for an homologous system hosted on a symmetric support. To argue along this line, consider a symmetric network of couplings: the spectrum of the Laplacian matrix is now real ($\Lambda_{Im}^{(\alpha)}=0$) and should hence fall on the real axis of Fig. \ref{fig:a155}(a). The grey region where the instability materializes, intersects the real axis only in the origin, where the trivial zero eigenvalue (corresponding to the uniform, fully synchronized, limit cycle state) sits. For the chosen values of $A$ and $B$, hence $c_1$ and $c_2$, it is therefore necessary to activate an imaginary component of  $\Lambda_{Im}^{(\alpha)}$, to invade the region of the complex plane deputed  to the instability. To state it differently, for the same choice of parameters, the instability sets in only if a suitable degree of asymmetry is enforced in the networks of connections, while it is bound to fade away otherwise. This  scenario is confirmed by Fig. \ref{fig:a155}(b) where the (magenta online) triangles  represent the real part of the dispersion relation, $\lambda_{Re}$, as a function of $-\Lambda_{Re}^{(\alpha)}$. Also in this case, the dispersion relation is punctually positive in correspondence of two specific entries $-\Lambda_{Re}^{(\alpha)}$. The solid line curve represents the homologous dispersion relation when the diffusive coupling is instead declined on a symmetric (and continuous) support.  

Patterns emerging  from the aforementioned instability are displayed in Fig. \ref{fig:a155}. Panel (c)  shows patterns relative to species $\phi$ obtained via a numerical integration of system \eqref{eq:RDbrusselator}, while panel (d) originates from the CGLE \eqref{eq:GLE}. The degree of correspondence is satisfactory and testifies a posteriori on the adequacy of the proposed approximation. 

\begin{figure*}[tb]
\begin{tabular}{ccc}
\includegraphics[scale=0.22]{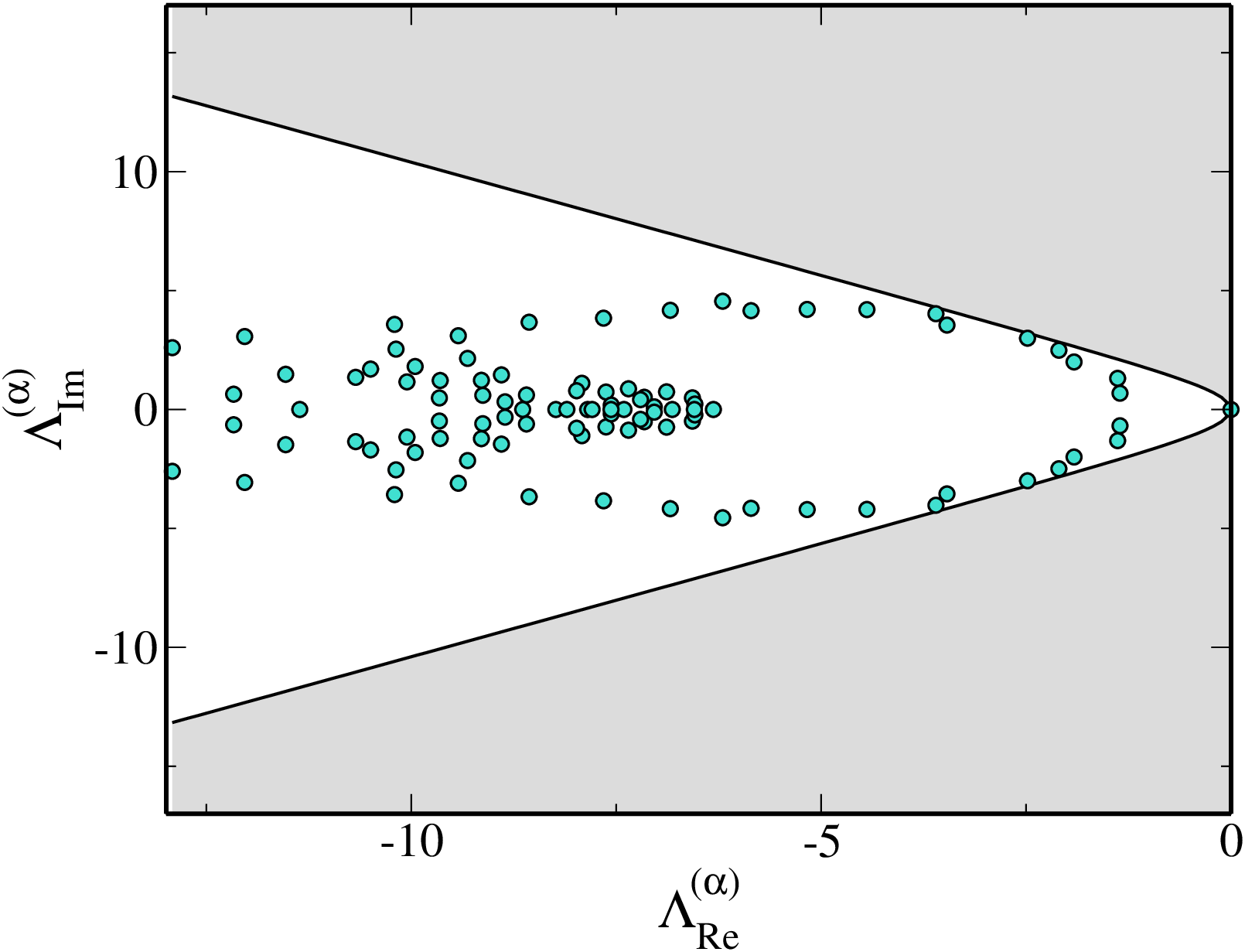} &
 \phantom{c} &
\includegraphics[scale=0.22]{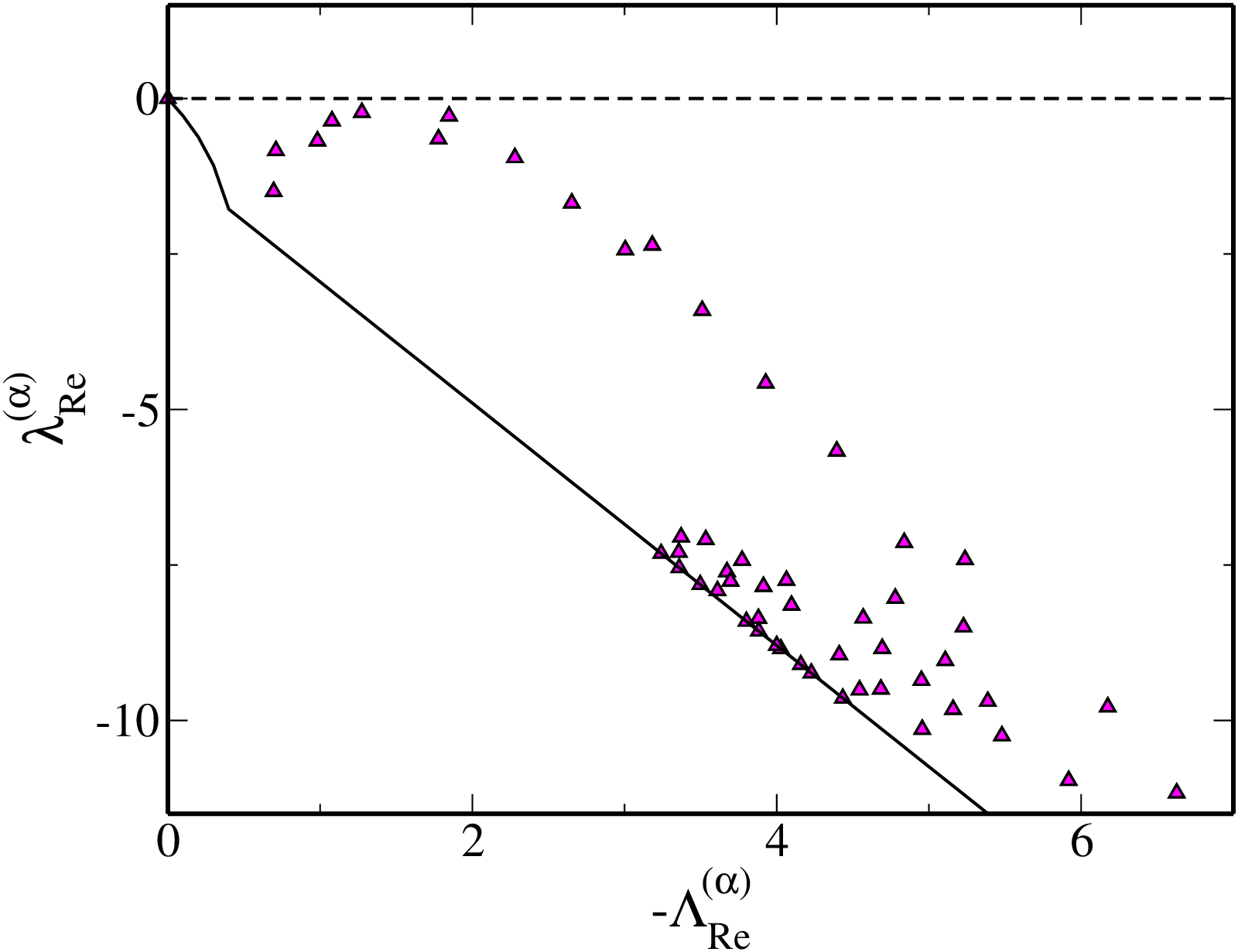}\\
(a) &  \phantom{c} & (b) \\
 \phantom{c} &  \phantom{c}\\
\includegraphics[scale=0.31]{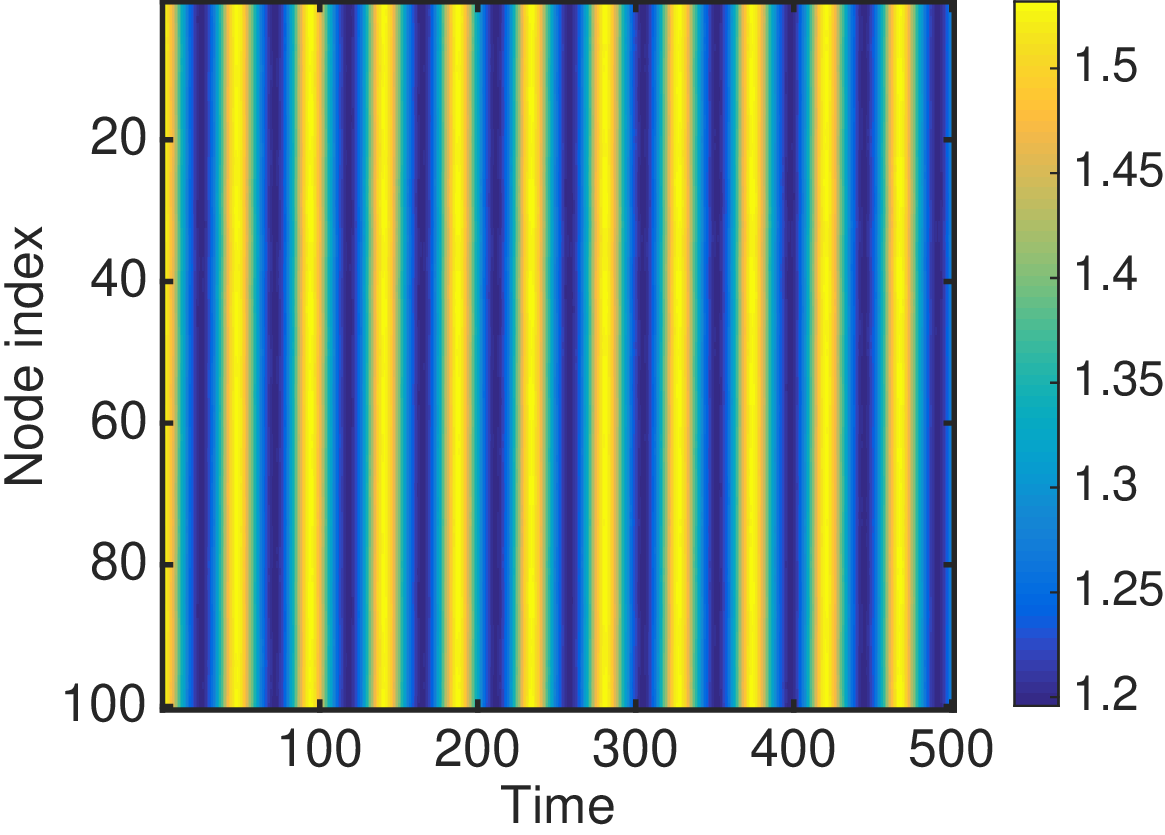} &
 \phantom{c} &
\includegraphics[scale=0.31]{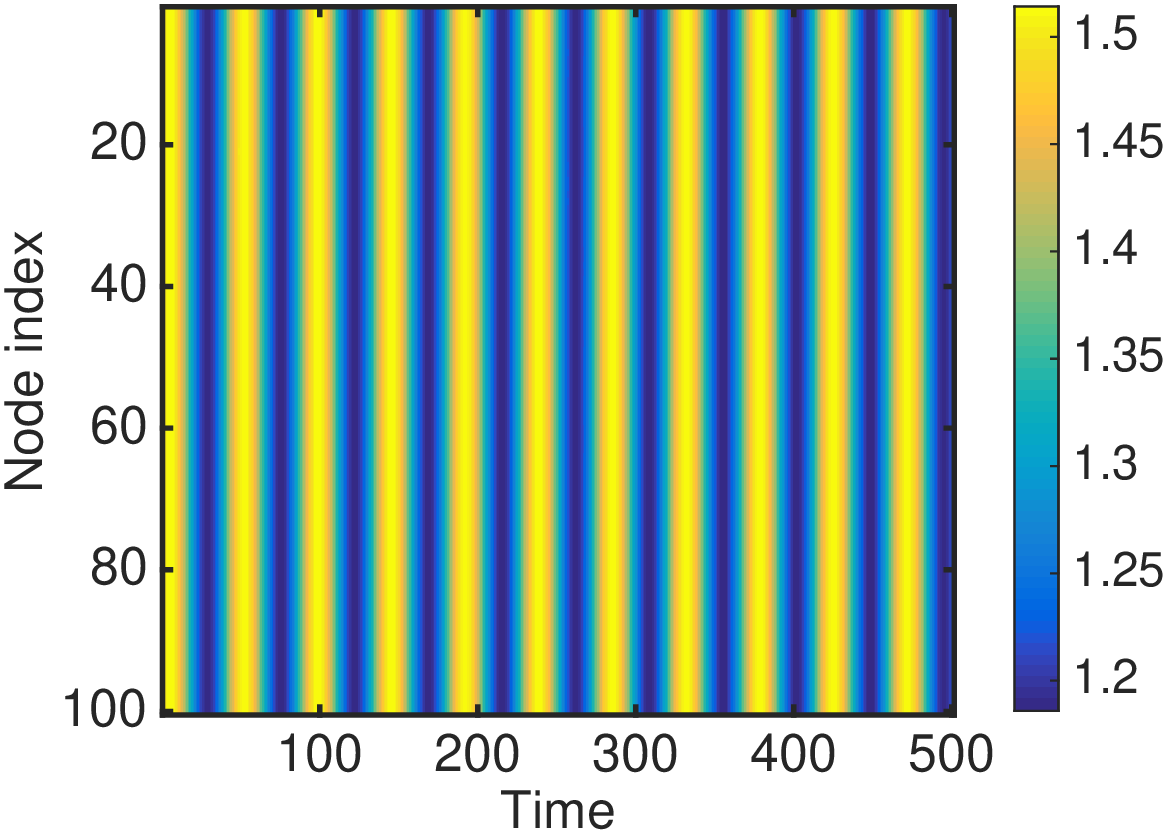}\\
(c) &  \phantom{c} & (d)
\end{tabular}
\caption{Panel (a): eigenvalues of the Laplacian for a balanced Newman-Watts \cite{newman99} network generated with $p=0.27$ and $N=100$ nodes. The algorithm to built the network is detailed in \cite{contemori}. The shaded area marks the instability region for the Brusselator model with $A=1.35<A_c$, $D_{\phi}=0.6$ and $D_{\psi}=4.9$.  The perturbative parameter is  $\epsilon=0.07$. Panel (b): the real part of the dispersion relation (magenta triangles) for the same choice of network and parameters as in panel (a). The black line originates from the continuous theory. Panel (c): pattern relative to species $\phi$ obtained by numerical integration of system \eqref{eq:RDbrusselator}. Panel (d): pattern relative to species $\phi$  calculated using the CGL equation.  \label{fig:a135}}
\end{figure*}

Figure \ref{fig:a135} refers to $A<A_c$: the synchronized limit cycle solution is predicted to be stable, following the CGL approximation. Direct integration of the original reaction diffusion system confirms this conclusion. The degree of correspondence between original and approximated schemes is, also in this case,  satisfying.
\section{Conclusions}
Synchronization is a fascinating field of investigation that has applications ranging from biology to computer science, just to name few examples. When this phenomenon occurs on complex networks, the stability of the  synchronized states  is  strongly affected by the topology of the graph. When the network is directed, external non homogeneous perturbations can destabilize synchronous oscillations also if the parameters are set to values that fall in the region of stability for the dynamics formulated on a symmetric support. 

Starting from these premises, we here considered  a reaction diffusion system defined on a directed (and balanced) graph. The system is made to evolve in the vicinity of a supercritical Hopf bifurcation and weak diffusive couplings are assumed to hold between adjacent nodes. By generalizing the work of Nakao \cite{nakao2014} to the setting where asymmetry in the couplings is enforced, and adapting the multiple-scales approach discussed in  Kuramoto \cite{kuramotoBook} to the case of a discrete, network like support, we derived a normal form equation to approximate the local evolution of the underlying reaction-diffusion system. This is a Complex Ginzburg-Landau equation (CGLE) whose coefficients clearly depend on the parameters of the model, but also on the topological characteristics of the hosting network. The CGLE can be effectively employed to probe the stability of the periodic uniform solution and work out the conditions for the onset of the symmetry breaking instability that eventually destroy the synchronized regime.  Numerical tests performed for the Brusselator model, confirm the adequacy of the proposed approximation. Patterns obtained under the CGLE scheme resemble closely those displayed upon integration of the original reaction-diffusion system.
\section{Acknowledgments}
This work has been supported by the program PRIN 2012 founded by the Italian Ministero dell'Istruzione, della Universit\`{a} e della Ricerca (MIUR). D.F. acknowledges financial support from H2020-MSCA-ITN-2015 project COSMOS  642563.  The work of T.C. presents research results of the Belgian Network DYSCO (Dynamical Systems, Control, and Optimization), funded by the Interuniversity Attraction Poles Programme, initiated by the Belgian State, Science Policy Office.
\appendix
\section{The solvability condition}\label{sec:solvability}
We consider  a linear operator $\bf A$, and two complex vectors ${\bf u}(t)$ and ${\bf b} (t)$ of the same length.  According to the Fredholm theorem, the  linear system ${ \bf A u}(t)={\bf b}(t)$ is solvable if $ \langle {\bf v} (t), {\bf b} (t) \rangle=0$ for all vectors ${\bf v}(t)$ solution of  ${\bf A}^{*} {\bf v}(t) =0$, where ${\bf A}^{*}$ is the adjoint operator satisfying  $\langle {\bf A} ^* \bf y, x \rangle = \langle  y,  A x \rangle$ $\forall x,y$. The angular brackets denote the scalar product that we here define as $\langle { \bf v}(t), {\bf b}(t) \rangle $ $ =\int_0^{2 \pi / \omega_0} {\bf v}^{\dagger}(t) {\bf b}(t) dt$, the symbol $\dagger$ standing for the conjugate transpose. 

With reference to equation \eqref{eq:terms}, the first requirement of the Fredholm theorem consists in finding ${\bf v}(t)$ such that  $(\partial / \partial t \mathbb{I}_{2 } -{\bf L}_0 )^{*} {\bf v}(t)=0$. By partial integration, and recalling that ${\bf L}_0$ is a real matrix, we find that $(\partial / \partial t \mathbb{I}_{2 } -{\bf L}_0 )^{*} =-(\partial / \partial t \mathbb{I}_{2 N} +{\bf L}_0)^{T}$. As a consequence, the system to be solved is $-(\partial / \partial t \mathbb{I}_{2} +{\bf L}_0)^{T} {\bf v}(t)=0$. In analogy with equation \eqref{eq:u1_W}, we search ${\bf v}(t)$ in the form ${\bf v} (t)={\bf U}_0^* e^{i \omega_0 t}$ for some vector ${\bf U}_0^*$. Substituting this ansatz into the previous equation, we find  ${\bf L}_0^T {\bf U}_0^*=-i \omega_0 {\bf U}_0^*$. To simplify calculation, it is convenient to normalize ${\bf U}_0^*$ so that $( {\bf U}_0^* )^{\dagger} {\bf U}_0= 1$.

Having defined ${\bf U}_0^*$, one can explicitly write down the  solvability condition $\langle {\bf U}_0^* e^{i \omega_0 t} , {\bf B}_j^{(\nu)} (t, \tau) \rangle =0$. Since $ {\bf B}_j^{(\nu)} (t, \tau)$ turns out to be periodic functions of period $2 \pi /\omega_0$, it is appropriate to express them in the form $ {\bf B}_j^{(\nu)} (t, \tau) =\sum_{l=-\infty}^{+\infty} \left  ( {\bf B}_j^{(\nu)} (\tau) \right )_l e^{i l \omega_0 t}$. If we multiply this series by  $({\bf U}_0^* e^{i \omega_0 t})^{\dagger}$ we again obtain periodic functions that, when integrated over the period $2\pi /\omega_0$ give zero. The only exception holds for $l=1$ which gives $\langle {\bf U}_0^* e^{i \omega_0 t} ,  \left ( {\bf B}_j^{(\nu)} (\tau) \right )_1  e^{i  \omega_0 t} \rangle = \int_0^{2 \pi / \omega_0} ({\bf U}_0^*)^{\dagger}  \left ( {\bf B}_j^{(\nu)} (\tau) \right )_1 dt$. The integrand does not depend on time $t$ and therefore the integral is zero only if the integrand itself  is identically equal to zero. For this reason the solvability condition reduces to $ ({\bf U}_0^*)^{\dagger}  \left ( {\bf B}_j^{(\nu)} (\tau) \right )_1 =0 $ $\forall j $.
\section{Details of the coefficients of the GL for the Brusselator model}\label{sec:detailsBrusselator}
In this section we explicitly derive the coefficients of the CGL equation for the Brusselator model.

The Jacobian matrix associate to system \eqref{eq:RDbrusselator} evaluated at the equilibrium point reads
\begin{equation*}
{\bf L}_0=
\begin{pmatrix} 
B-1 & A^2 \\ 
-B & -A^2  
\end{pmatrix} \quad .
\end{equation*}
Setting the system at the bifurcation point means  requiring ${\bf L}_0$ to have a pair of imaginary eigenvalues, which translates into $\text{tr}{\bf L}_0=0$. This relation implies a constrain on the critical value of the parameter $B$ which must be chosen as $B_{c}=1+A^2$. With this assumption, ${\bf L}_0$ can be rewritten as   
\begin{equation*}
{\bf L}_0=
\begin{pmatrix} 
A^2 & A^2 \\ 
-(1+A^2) & -A^2  
\end{pmatrix}
\end{equation*} 
and its pair of imaginary eigenvalues are  $\lambda_0=\pm i\omega_0=\pm iA$.  The right eigenvectors of  $L_0$ and  $L_0^{T}$ corresponding to, respectively, eigenvalues $i A$ and $-i A$ are given by
\begin{equation*}
\begin{aligned}
 {\bf U}_0 &=
\begin{pmatrix}
1\\
-1+\dfrac{i}{A}
\end{pmatrix} \\
 {\bf U}_0^{*} &=\dfrac{1}{2}
\begin{pmatrix}
1+iA \\
iA
\end{pmatrix}  \quad . 
\end{aligned}
\end{equation*}

To proceed with the multiscale analysis, we perturb $B_c$ as $B=B_c+\epsilon^2B_c$ and we immediately find the entries of matrix $ {\bf L}_1$ 
\begin{equation*}
 {\bf L}_1=(1+A^2)
 \begin{pmatrix}
1 & 0 \\ 
-1 & 0
\end{pmatrix}
\end{equation*}
and the value of $\sigma$ which turns out to be $\dfrac{(1+A^2)}{2} $. 

The operators  $\mathcal{M}_0$ and $\mathcal{N}_0$ are given  by 
\begin{equation*}
\begin{aligned}
 ( \mathcal{M}_0 {\bf x y })_1 & = -( \mathcal{M}_0 {\bf x y })_2\\
 &=\frac{1+A^2}{A} x_1 y_1 +A(x_1 y_2+x_2  y_1)  \\
 ( \mathcal{N}_0 {\bf x y z })_1&=  - ( \mathcal{N}_0 {\bf x y z })_2 \\
 &=\dfrac{1}{3}(x_1 y_1 z_2 +x_2 y_1 z_1 +x_1 y_2 z_1)
\end{aligned}
\end{equation*} 
where $(\cdot)_i$ defines the $i$-th component of a two-dimensional vector. These preliminary quantities allow us to calculate
\begin{equation}\label{eq:M0}
\begin{aligned}
\mathcal{M}_0 {\bf U}_0 {\bf U}_0  &=
\begin{pmatrix}
\dfrac{1}{A}-A+2i \\
\phantom{i}\\
 -\dfrac{1}{A}+A-2i
\end{pmatrix}\\
\mathcal{M}_0 {\bf U}_0 \bar{ {\bf U}}_{0}  &=
\begin{pmatrix}
\dfrac{1}{A} -A \\
\phantom{i}\\
-\dfrac{1}{A} +A
\end{pmatrix}\\
\end{aligned}
\end{equation}
and
\begin{equation*}
\mathcal{N}_0 {\bf U}_0 {\bf U}_0 \bar{ {\bf U}}_{0} =
\begin{pmatrix}
-1+\dfrac{i}{3 A} 	\\
\phantom{i}\\
1 -\dfrac{i}{3 A}
\end{pmatrix} \quad .
\end{equation*}
From Eq. \eqref{eq:M0} we can make explicit $ {\bf V}_0$ and  ${\bf V}_2$
\begin{equation*}
\begin{aligned}
 {\bf V}_0 &=\dfrac{2(A^2-1)}{A^3}
 \begin{pmatrix}
 0\\ 1 
 \end{pmatrix}\\
{\bf V}_2 &=-\dfrac{1}{3}
\begin{pmatrix}
-\dfrac{4}{A}+2 \left (\dfrac{1}{A^2}-1 \right )i\\
\phantom{i}\\
-\dfrac{1}{A} \left [ \left (\dfrac{1}{A^2}-5 \right )+ 2  \left ( \dfrac{2}{A}-A \right ) i  \right ]
\end{pmatrix}
\end{aligned}
\end{equation*}
and thus 
\begin{equation*}
\begin{aligned}
\mathcal{M}_0\bar { { \bf U}}_{0} {\bf V}_{2} & =
\begin{pmatrix}
-1+\dfrac{1}{A^2} +i\dfrac{2}{3}\left [ \dfrac{A^2 -A^4 -1}{A^3} \right ] \\
\phantom{i}\\
 +1-\dfrac{1}{A^2} -i\dfrac{2}{3} \left [ \dfrac{A^2 -A^4 -1}{A^3} \right ]
 \end{pmatrix}
\\
\mathcal{M}_0 {\bf U}_0 {\bf V}_0 & =
\begin{pmatrix}
\dfrac{2(A^2-1)}{A^2} \\
\phantom{i}\\
-\dfrac{2(A^2-1)}{A^2}
\end{pmatrix} \quad .
\end{aligned}
\end{equation*}
Notice that our equations for ${ \bf U}_{0} $ and ${\bf V}_{2} $ are slightly different from those reported in \cite{kuramotoBook}. Despite this minor difference, we end up with same values of $d$ and $g$. Finally, we can write down the two last  coefficients of the CGL equation:
\begin{eqnarray*}
d&=&\dfrac{1}{2} \left [ D_u +D_v -iA \left ( D_u-D_v \right ) \right ]\\
g&=&\dfrac{1}{2}\left [ \dfrac{A^2+2}{A^2}+i\left ( \dfrac{4A^4+4-7A^2}{3A^3}\right ) \right ] \quad .
\end{eqnarray*}
\bibliographystyle{model1-num-names}
\bibliography{bibliography.bib}
\end{document}